\documentclass[aps,twocolumn]{revtex4}

\usepackage{graphics}
\usepackage{epsfig}
\usepackage{bm}
\usepackage{amsmath,amssymb,framed,color}
\usepackage{wasysym}

\def\vb#1{\mbox{\boldmath$#1$}}
\def\pd#1#2{\frac{\partial #1}{\partial #2}}
\def\fd#1#2{\frac{\delta #1}{\delta #2}}
\def\wh#1{\widehat{#1}}
\def\bdot{\,\vb{\cdot}\,}
\def\btimes{\,\vb{\times}\,}

\def\bhat{\wh{{\sf b}}}
\def\cal#1{\mathcal{#1}}

\def\bhat{\wh{{\sf b}}}

\begin{document}

\title{Hamiltonian structure of the guiding-center Vlasov-Maxwell equations}

\author{Alain J.~Brizard}
\affiliation{Department of Physics, Saint Michael's College, Colchester, VT 05439, USA}

\begin{abstract}
The Hamiltonian structure of the guiding-center Vlasov-Maxwell equations is presented in terms of a Hamiltonian functional and a guiding-center Vlasov-Maxwell bracket. The bracket, which is shown to satisfy the Jacobi identity exactly, is used to show that the guiding-center momentum and angular-momentum conservation laws can also be expressed in Hamiltonian form. 
\end{abstract}

\date{\today}


\maketitle

\section{Introduction}

The complementary Lagrangian and Hamiltonian formulations of dissipationless plasma equations have a rich history in plasma physics \cite{Morrison_2005,Morrison_2017}. In the Lagrangian formulation for the Vlasov-Maxwell equations, for example, one of several types of variational (Euler, Lagrange, or Euler-Poincar\'{e}) principles is used to derive these dissipationless plasma equations \cite{Ye_Morrison_1992,Brizard_2000}, while their exact conservation laws can be derived through the Noether method.

In the Hamiltonian formulation of dissipationless plasma equations, on the other hand, the equations for a given set of plasma fields $\vb{\psi}$ are, first, written in Hamiltonian form: 
\begin{equation}
\pd{\psi^{a}}{t} \;=\; {\sf J}^{ab}\circ\fd{\cal H}{\psi^{b}}, 
\label{eq:Ham_psi}
\end{equation}
in terms of a Hamiltonian functional ${\cal H}[\vb{\psi}]$ and the antisymmetric Poisson operator ${\sf J}^{ab}(\vb{\psi})$ acting (denoted as $\circ$) on the functional derivative $\delta{\cal H}/\delta\psi^{b}$. Next, the evolution of an arbitrary functional ${\cal F}[\vb{\psi}]$ is expressed in Hamiltonian form
\begin{equation}
\pd{\cal F}{t} = \left\langle \fd{\cal F}{\psi^{a}}\left|\frac{}{}\right.\pd{\psi^{a}}{t} \right\rangle = \left\langle \fd{\cal F}{\psi^{a}}\left|\frac{}{}\right. {\sf J}^{ab}\circ\fd{\cal H}{\psi^{b}} \right\rangle \equiv \left[{\cal F},\frac{}{} {\cal H}\right],
\label{eq:F_Ham}
\end{equation}
which is then used to construct a bracket $[\;,\;]$ on functionals, where the inner product $\langle\;|\;\rangle$ involves suitable integrations over the domains of the field components.

For example, we consider the Vlasov-Maxwell equations with field components $\vb{\psi} = (f,{\bf E},{\bf B})$, which are expressed in Hamiltonian form \eqref{eq:Ham_psi}:
\begin{eqnarray}
\pd{f({\bf x},{\bf p},t)}{t} &=& -\;\left\{ f,\; \fd{\cal H}{f}\right\} \;+\; 4\pi q\,\left\{ f,\; {\bf x}\right\}\bdot\fd{\cal H}{\bf E}\ \nonumber \\
 &\equiv& {\sf J}^{ff}\circ\fd{\cal H}{f} + {\sf J}^{f{\bf E}}\circ\fd{\cal H}{\bf E}, \label{eq:Vlasov_Ham} \\
\pd{{\bf E}({\bf x},t)}{t} & = & 4\pi c\,\nabla\btimes\fd{\cal H}{\bf B} \;+\; 4\pi q \int_{\bf p} \left\{ {\bf x},\; f\right\}\,\fd{\cal H}{f} \nonumber \\
 &\equiv& {\sf J}^{{\bf E}{\bf B}}\circ\fd{\cal H}{\bf B} + {\sf J}^{{\bf E}f}\circ\fd{\cal H}{f}, \label{eq:Maxwell_Ham} \\
\pd{{\bf B}({\bf x},t)}{t} & = & -\,4\pi c\,\nabla\btimes\fd{\cal H}{\bf E} \equiv {\sf J}^{{\bf B}{\bf E}}\circ\fd{\cal H}{\bf E}, \label{eq:Faraday_Ham}
\end{eqnarray}
where the Vlasov function $f$ is defined in six-dimensional particle phase space ${\bf z} = ({\bf x},{\bf p})$, while the electric and magnetic fields $({\bf E},{\bf B})$ are defined in three-dimensional configuration space ${\bf x}$.
with the single-particle noncanonical Poisson bracket
\begin{equation}
\{f,\; g\} \;=\; \nabla f\bdot\pd{g}{\bf p} - \pd{f}{\bf p}\bdot\nabla g \;+\; \frac{q}{c}{\bf B}\bdot\pd{f}{\bf p}\btimes\pd{g}{\bf p},
\label{eq:Poisson_xp}
\end{equation}
and the Hamiltonian functional
\begin{equation}
{\cal H} \;=\; \int_{\bf z} f\;\frac{|{\bf p}|^{2}}{2m} \;+\; \int_{\bf x} \frac{1}{8\pi} \left(|{\bf E}|^{2} \;+\frac{}{} |{\bf B}|^{2} \right),
\label{eq:VM_Ham}
\end{equation}
with $\delta{\cal H}/\delta f = |{\bf p}|^{2}/2m$, $\delta{\cal H}/\delta{\bf E} = {\bf E}/4\pi$, and $\delta{\cal H}/\delta{\bf B} = {\bf B}/4\pi$. In Eqs.~\eqref{eq:Maxwell_Ham} and \eqref{eq:VM_Ham}, summation over particle species is implied whenever an integration over the Vlasov distribution $f$ appears, where $\int_{\bf p}$ and $\int_{\bf z}$ denote integrations over particle momentum and particle phase space, respectively, while 
$\int_{\bf x}$ denotes an integral over configuration space.

By expressing the evolution of an arbitrary functional ${\cal F}[f,{\bf E},{\bf B}]$ on the Vlasov-Maxwell fields
\begin{eqnarray}
\pd{\cal F}{t} &=& \int_{\bf z}\pd{f}{t}\;\fd{\cal F}{f} \;+\; \int_{\bf x} \left(\pd{\bf E}{t}\bdot\fd{\cal F}{\bf E} \;+\; \pd{\bf B}{t}\bdot\fd{\cal F}{\bf B}\right) \nonumber \\
 &\equiv& \left\langle \fd{\cal F}{\psi^{a}}\left|\frac{}{}\right.\pd{\psi^{a}}{t} \right\rangle =  \left\langle \fd{\cal F}{\psi^{a}}\left|\frac{}{}\right. {\sf J}^{ab}\circ\fd{\cal H}{\psi^{b}} \right\rangle,
\end{eqnarray}
we easily arrive at the Vlasov-Maxwell bracket \cite{M,MW,B}
\begin{eqnarray}
\pd{\cal F}{t} &=& \int_{\bf z} f\;\left\{ \fd{\cal F}{f},\; \fd{\cal H}{f}\right\} \nonumber \\
 &&+\; 4\pi q  \int_{\bf z} f \left(\fd{\cal H}{\bf E}\bdot\left\{{\bf x}, \fd{\cal F}{f}\right\} - \fd{\cal F}{\bf E}\bdot\left\{{\bf x}, \fd{\cal H}{f}\right\} \right) \nonumber \\
 &&+\; 4\pi c \int_{\bf x}\left( \fd{\cal F}{\bf E}\bdot\nabla\btimes\fd{\cal H}{\bf B} \;-\; \fd{\cal H}{\bf E}\bdot\nabla\btimes\fd{\cal F}{\bf B}\right) \nonumber \\
  &\equiv& \left[{\cal F},\frac{}{} {\cal H}\right],
 \label{eq:VM_bracket}
 \end{eqnarray}
 where integrations by parts were performed. The proof that the Vlasov-Maxwell bracket \eqref{eq:VM_bracket} satisfies the Jacobi identity for arbitrary functionals $({\cal F},{\cal G},{\cal K})$:
 \begin{equation}
 \left[[{\cal F},\,{\cal G}],\frac{}{} {\cal K}\right] +  \left[[{\cal G},\,{\cal K}],\frac{}{} {\cal F}\right] +  \left[[{\cal K},\,{\cal F}],\frac{}{} {\cal G}\right] = 0,
 \end{equation}
was given in the Appendix of Ref.~\cite{Morrison_2013}, where the Poisson bracket \eqref{eq:Poisson_xp} was broken into canonical and non-canonical parts, and Appendix B of Ref.~\cite{Brizard_2016_arxiv}, where properties of the full Poisson bracket \eqref{eq:Poisson_xp} were used. We note that the Jacobi property of the Vlasov-Maxwell bracket \eqref{eq:VM_bracket} is  inherited from the Jacobi property of the Poisson bracket \eqref{eq:Poisson_xp}, which requires that $\nabla\bdot{\bf B} = 0$.
 
 The purpose of the present paper is to follow a similar construction for the guiding-center Hamiltonian structure directly from the guiding-center Vlasov-Maxwell equations. This approach is in contrast to the Lie-transform construction of a Hamiltonian structure for the reduced Vlasov-Maxwell equations \cite{Brizard_2016}, which is derived from the Vlasov-Maxwell bracket \eqref{eq:VM_bracket} and automatically guarantees that the reduced Vlasov-Maxwell bracket satisfies the Jacobi property. Here, while there is no guarantee that the guiding-center Vlasov-Maxwell bracket will satisfy the Jacobi property, its derivation is simple.

\section{Guiding-center Vlasov-Maxwell Equations}

The variational formulations of the guiding-center Vlasov-Maxwell equations were presented by Pfirsch and Morrison \cite{Pfirsch_Morrison_1985} and more recently by Brizard and Tronci \cite{Brizard_Tronci_2016}, whose works also included a derivation of exact conservation laws for energy-momentum and angular momentum through the Noether method. The guiding-center equations of motion considered here are the simplest equations derived from a variational principle \cite{Littlejohn_1983,Cary_Brizard_2009}. 

First, the guiding-center single-particle Lagrangian for a charged particle (of charge $q$ and mass $m$) moving in a reduced phase space, with guiding-center position ${\bf X}$ and guiding-center parallel momentum $p_{\|}$, is expressed as
\begin{equation}
L_{\rm gc} \;=\; \left(\frac{q}{c}\,{\bf A} + p_{\|}\,\bhat\right)\bdot\frac{d{\bf X}}{dt} \;-\; \left(q\,\Phi \;+\frac{}{} K_{\rm gc}\right),
\label{eq:Lag_gc}
\end{equation}
where the electromagnetic potentials $(\Phi,{\bf A})$ yield the electric field ${\bf E} = -\,\nabla\Phi - c^{-1}\partial{\bf A}/\partial t$ and the magnetic field ${\bf B} = \nabla\btimes{\bf A} \equiv B\,\bhat$, and the guiding-center kinetic energy is $K_{\rm gc} = 
p_{\|}^{2}/2m + \mu\,B$, where $\mu$ denotes the guiding-center magnetic moment (which is a guiding-center invariant). The guiding-center equations of motion are derived from the guiding-center Lagrangian \eqref{eq:Lag_gc} as Euler-Lagrange equations, which are expressed as
\begin{eqnarray}
\frac{d{\bf X}}{dt} &=& \left\{{\bf X},\; K_{\rm gc}\right\}_{\rm gc} \;+\; q\,{\bf E}^{*}\bdot\{{\bf X},{\bf X}\}_{\rm gc}, \label{eq:X_dot_gc} \\
\frac{dp_{\|}}{dt} &=& \left\{p_{\|},\; K_{\rm gc}\right\}_{\rm gc} \;+\; q\,{\bf E}^{*}\bdot\{{\bf X}, p_{\|}\}_{\rm gc}, \label{eq:p_dot_gc}  
 \end{eqnarray}
 where the guiding-center Poisson bracket \cite{Cary_Brizard_2009}
\begin{eqnarray}
\{ f,\; g\}_{\rm gc} &\equiv& \frac{{\bf B}^{*}}{B_{\|}^{*}}\bdot\left(\nabla f\;\pd{g}{p_{\|}} \;-\; \pd{f}{p_{\|}}\;\nabla g\right) \nonumber \\
 &&-\; \frac{c\bhat}{qB_{\|}^{*}}\bdot\nabla f\btimes\nabla g
\label{eq:PB_gc}
\end{eqnarray}
is used without the ignorable gyromotion pair $(\mu,\theta)$, and the effective fields are 
\begin{equation}
\left. \begin{array}{rcl}
{\bf E}^{*} &\equiv & {\bf E} \;-\; (p_{\|}/q)\,\partial\bhat/\partial t \\
{\bf B}^{*} & \equiv & {\bf B} \;+\; (p_{\|}c/q)\,\nabla\btimes\bhat \\
B_{\|}^{*} & \equiv & \bhat\bdot{\bf B}^{*} \;=\; B \;+\; (p_{\|}c/q)\,\bhat\bdot\nabla\btimes\bhat
\end{array} \right\}.
\label{eq:B*_def}
\end{equation}
We note that the guiding-center Poisson bracket \eqref{eq:PB_gc} satisfies the Jacobi property for arbitrary functions $(f,g,h)$:
\begin{equation}
\left\{ \{f, g\}_{\rm gc},\frac{}{} h\right\}_{\rm gc} + \left\{ \{g, h\}_{\rm gc},\frac{}{} f\right\}_{\rm gc} + \left\{ \{h, f\}_{\rm gc},\frac{}{} g\right\}_{\rm gc} = 0,
\label{eq:gcPB_Jacobi}
\end{equation}
subject to the condition
\begin{equation}
\nabla\bdot{\bf B}^{*} \;=\; \nabla\bdot{\bf B} \;=\; 0,
\label{eq:div_Bstar}
\end{equation}
which is satisfied by the definition \eqref{eq:B*_def}. We also note that the guiding-center Jacobian ${\cal J}_{\rm gc} \equiv 2\pi\,m\,B_{\|}^{*}$ satisfies the guiding-center Liouville theorem
\begin{equation}
\pd{{\cal J}_{\rm gc}}{t} + \nabla\bdot\left({\cal J}_{\rm gc}\,\frac{d{\bf X}}{dt}\right) + \pd{}{p_{\|}}\left({\cal J}_{\rm gc}\,\frac{dp_{\|}}{dt}\right) = 0,
\label{eq:Liouville_gc}
\end{equation}
with the guiding-center equations of motion \eqref{eq:X_dot_gc}-\eqref{eq:p_dot_gc}.

Next, we introduce the guiding-center Vlasov-Maxwell equations \cite{Brizard_Tronci_2016} for the guiding-center fields $\vb{\Psi}_{\rm gc} = (F_{\rm gc}, {\bf E},{\bf B})$:
\begin{eqnarray}
\pd{F_{\rm gc}}{t}  &=& -\,\nabla\bdot\left(F_{\rm gc}\,\frac{d{\bf X}}{dt}\right) - \pd{}{p_{\|}}\left(F_{\rm gc}\,\frac{dp_{\|}}{dt}\right) , \label{eq:V_eq} \\
\pd{\bf E}{t} &=& c\,\nabla\btimes\left({\bf B} -\frac{}{} 4\pi\,{\bf M}_{\rm gc}\right) - 4\pi q\int_{P} F_{\rm gc}\,\frac{d{\bf X}}{dt} \nonumber \\
 &\equiv& c\,\nabla\btimes{\bf H}_{\rm gc} \;-\; 4\pi\,{\bf J}_{\rm gc}, \label{eq:Maxwell_eq} \\
\pd{\bf B}{t} &=& -\,c\,\nabla\btimes{\bf E}, \label{eq:Faraday_eq} 
\end{eqnarray}
where the phase-space density $F_{\rm gc} \equiv F\,{\cal J}_{\rm gc}$ is defined in terms of the guiding-center Vlasov function $F$ and the guiding-center Jacobian ${\cal J}_{\rm gc}$, the guiding-center momentum integral $\int_{P} \equiv \int d p_{\|}\,d\mu$ excludes the guiding-center Jacobian ${\cal J}_{\rm gc}$, and summation over particle species is implied whenever an integral over $F_{\rm gc}$ appears. In addition, the guiding-center magnetic field ${\bf H}_{\rm gc} \equiv {\bf B} - 4\pi\,{\bf M}_{\rm gc}$ is defined in terms of the guiding-center magnetization
\begin{equation}
{\bf M}_{\rm gc} \;\equiv\; \int_{P} F_{\rm gc}\left(-\; \mu\,\bhat \;+\; \frac{q}{c}\,\mathbb{P}_{\|}\bdot\frac{d{\bf X}}{dt} \right),
\label{eq:M_def}
\end{equation}
which is expressed in terms of the intrinsic guiding-center magnetization $-\,\mu\,\bhat$ and the moving guiding-center electric-dipole moment \cite{Tronko_Brizard_2015}
 \begin{equation}
\left( \frac{q\bhat}{\Omega}\btimes\frac{d{\bf X}}{dt}\right)\btimes\frac{p_{\|}\bhat}{mc} \;\equiv\;  \frac{q}{c}\,\mathbb{P}_{\|}\bdot\frac{d{\bf X}}{dt},
 \end{equation}
where we introduced the symmetric dyadic tensor
\begin{equation}
 \mathbb{P}_{\|} \;\equiv\; \frac{cp_{\|}}{qB}\;\left({\bf I} \;-\; \bhat\,\bhat\right).
 \label{eq:P_par}
 \end{equation}
We note that, while the guiding-center magnetization \eqref{eq:M_def} is derived from the guiding-center Lagrangian \eqref{eq:Lag_gc}: ${\bf M}_{\rm gc} \equiv \int_{P} F_{\rm gc}\,\delta L_{\rm gc}/\delta{\bf B}$, the guiding-center polarization 
${\bf P}_{\rm gc} \equiv \int_{P} F_{\rm gc}\,\delta L_{\rm gc}/\delta{\bf E} \equiv 0$ is absent in our model \cite{Footnote}.

We now express Eqs.~\eqref{eq:V_eq}-\eqref{eq:Faraday_eq} in Hamiltonian form $\partial\Psi_{\rm gc}^{a}/\partial t \equiv {\sf J}_{\rm gc}^{ab}\circ\delta{\cal H}_{\rm gc}/\delta\Psi_{\rm gc}^{b}$:
\begin{eqnarray}
\pd{F_{\rm gc}}{t} &=& -\;B_{\|}^{*}\left\{ \frac{F_{\rm gc}}{B_{\|}^{*}},\; \fd{{\cal H}_{\rm gc}}{F_{\rm gc}} \right\}_{\rm gc} \nonumber \\
 &&-\; 4\pi q \frac{\delta^{\star}{\cal H}_{\rm gc}}{\delta{\bf E}}\bdot B_{\|}^{*}\left\{ {\bf X},\; \frac{F_{\rm gc}}{B_{\|}^{*}}\right\}_{\rm gc},
\label{eq:V_bracket} \\
 \pd{\bf E}{t} & = & 4\pi c\,\nabla\btimes\left( \fd{{\cal H}_{\rm gc}}{\bf B} \;-\; \frac{q}{c}\;\int_{P} F_{\rm gc}\;\mathbb{P}_{\|}\bdot\frac{d{\bf X}}{dt} \right) \nonumber \\
  &&-\; 4\pi q \int_{P}F_{\rm gc}\;\frac{d{\bf X}}{dt},
  \label{eq:B_bracket} \\
\pd{\bf B}{t} &=& -\,4\pi c\;\nabla\btimes\fd{{\cal H}_{\rm gc}}{\bf E},
  \label{eq:E_bracket}
  \end{eqnarray}
where the guiding-center velocity in Eq.~\eqref{eq:B_bracket} is
 \begin{equation}
 \frac{d{\bf X}}{dt} \;=\;  \left\{ {\bf X},\; \fd{{\cal H}_{\rm gc}}{F_{\rm gc}} \right\}_{\rm gc} + 4\pi q\;\frac{\delta^{\star}{\cal H}_{\rm gc}}{\delta{\bf E}}\bdot\{{\bf X},\; {\bf X}\}_{\rm gc}.
 \end{equation}
Here, the guiding-center Hamiltonian functional is
\begin{equation}
{\cal H}_{\rm gc} \;\equiv\; \int_{\bf Z} F_{\rm gc}\;K_{\rm gc} \;+\;  \int_{\bf X} \frac{1}{8\pi}\left(|{\bf E}|^{2} + |{\bf B}|^{2}\right),
\label{eq:gc_H}
\end{equation}
where $\int_{\bf Z}$ denotes an integration over the guiding-center phase space while $\int_{\bf X}$ denotes an integral over the three-dimensional guiding-center configuration space, from which we obtain the functional derivatives
\begin{equation}
\left( \begin{array}{c}
\delta{\cal H}_{\rm gc}/\delta F_{\rm gc} \\
\delta{\cal H}_{\rm gc}/\delta{\bf E} \\
\delta{\cal H}_{\rm gc}/\delta{\bf B}
\end{array} \right) \;=\; \left( \begin{array}{c}
K_{\rm gc} \\
{\bf E}/4\pi \\
{\bf B}/4\pi + \int_{P}F_{\rm gc}\;\mu\,\bhat
\end{array} \right),
\label{eq:delta_H}
\end{equation}
where $\delta K_{\rm gc}/\delta{\bf B} = \mu\,\bhat$ and we introduced  the definition
 \begin{eqnarray}
4\pi q\,\frac{\delta^{\star}{\cal H}_{\rm gc}}{\delta{\bf E}} &\equiv& 4\pi q \left(\fd{{\cal H}_{\rm gc}}{\bf E} + \mathbb{P}_{\|}\bdot\nabla\btimes\fd{{\cal H}_{\rm gc}}{\bf E}\right) \nonumber \\
 &=&q\,{\bf E} \;-\; p_{\|}\,\pd{\bhat}{t} \;\equiv\; q\,{\bf E}^{*},
\label{eq:delta_star}
\end{eqnarray}
after making use of Faraday's law \eqref{eq:E_bracket}. We note that the additional guiding-center Maxwell equations are
\begin{eqnarray}
\nabla\bdot{\bf E} &=& 4\pi q\,\int_{P} F_{\rm gc} \;\equiv\; 4\pi\,\varrho_{\rm gc}, \label{eq:gc_divE} \\
\nabla\bdot{\bf B} &=& 0, \label{eq:gc_divB}
\end{eqnarray}
where Eq.~\eqref{eq:gc_divE} is connected to Eq.~\eqref{eq:Maxwell_eq} through the guiding-center charge conservation law $\partial\varrho_{\rm gc}/\partial t + \nabla\bdot{\bf J}_{\rm gc} = 0$, while Eq.~\eqref{eq:gc_divB} can be viewed as an initial condition of the Faraday equation \eqref{eq:Faraday_eq}.

\section{\label{sec:Jacobi}Guiding-center Vlasov-Maxwell bracket}

The guiding-center Vlasov-Maxwell bracket is now constructed from the guiding-center Vlasov-Maxwell equations \eqref{eq:V_bracket}-\eqref{eq:E_bracket} and the Hamiltonian functional \eqref{eq:gc_H}:
\begin{eqnarray}
\pd{\cal F}{t} &=& \int_{\bf Z}\pd{F_{\rm gc}}{t}\;\fd{\cal F}{F_{\rm gc}} +  \int_{\bf X}\left(\pd{\bf E}{t}\bdot\fd{\cal F}{\bf E} + \pd{\bf B}{t}\bdot\fd{\cal F}{\bf B}\right) \nonumber \\
  &\equiv& \left\langle \fd{\cal F}{\Psi^{a}} \left|\frac{}{}\right. \pd{\Psi^{a}}{t}\right\rangle = \left\langle \fd{\cal F}{\Psi^{a}}\left|\frac{}{}\right. {\sf J}_{\rm gc}^{ab}\circ\fd{{\cal H}_{\rm gc}}{\Psi^{b}} \right\rangle ,
\label{eq:F_fd}
\end{eqnarray} 
where the guiding-center Vlasov-Maxwell bracket for two arbitrary functionals $({\cal F},{\cal G})$ of the fields $\vb{\Psi} = (F_{\rm gc},{\bf E},{\bf B})$ is defined in terms of the Poisson structure:
\begin{equation}
\left[{\cal F},\frac{}{} {\cal G}\right]_{\rm gc} \;\equiv\; \left\langle \fd{\cal F}{\Psi^{a}}\;\left|\frac{}{}\right. {\sf J}_{\rm gc}^{ab}\circ\fd{\cal G}{\Psi^{b}} \right\rangle.
\label{eq:Poisson_structure}
\end{equation}
Here, the antisymmetric Poisson operator ${\sf J}_{\rm gc}^{ab}(\vb{\Psi})$ guarantees the antisymmetry property: $[{\cal F},{\cal G}]_{\rm gc} = -\,[{\cal G},{\cal F}]_{\rm gc}$; and the bilinearity of Eq.~\eqref{eq:Poisson_structure} guarantees the Leibniz property: $[{\cal F},{\cal G}\,{\cal K}]_{\rm gc} = [{\cal F},{\cal G}]_{\rm gc}\,{\cal K} + {\cal G}\,[{\cal F},{\cal K}]_{\rm gc}$. The Jacobi property:
\begin{eqnarray}
{\cal Jac}[{\cal F},{\cal G},{\cal K}] &\equiv& \left[[{\cal F},{\cal G}]_{\rm gc},\frac{}{} {\cal K}\right]_{\rm gc} + \left[[{\cal G},{\cal K}]_{\rm gc},\frac{}{} {\cal F}\right]_{\rm gc} \nonumber \\
 &&+ \left[[{\cal K},{\cal F}]_{\rm gc},\frac{}{} {\cal G}\right]_{\rm gc} \;=\; 0,
\label{eq:Jacobi}
\end{eqnarray}
which holds for arbitrary functionals $({\cal F},{\cal G},{\cal K})$, involves the guiding-center Poisson operator ${\sf J}_{\rm gc}^{ab}(\vb{\Psi})$. 

From Eq.~\eqref{eq:F_fd}, we can now extract the guiding-center Vlasov-Maxwell bracket expressed in terms of two arbitrary guiding-center functionals $({\cal F},{\cal G})$ as
 \begin{eqnarray}
 \left[{\cal F},\frac{}{}{\cal G}\right]_{\rm gc} &=& \int_{\bf Z} F_{\rm gc} \left\{ \fd{{\cal F}}{F_{\rm gc}} ,\; \fd{\cal G}{F_{\rm gc}} \right\}_{\rm gc} 
  \label{eq:gcVM_bracket} \\
  &&+\; 4\pi q \int_{\bf Z} F_{\rm gc}\;\frac{\delta^{\star}{\cal G}}{\delta{\bf E}}\bdot\left\{{\bf X},\; \fd{{\cal F}}{F_{\rm gc}} \right\}_{\rm gc} \nonumber \\
   &&-\; 4\pi q \int_{\bf Z} F_{\rm gc}\;\frac{\delta^{\star}{\cal F}}{\delta{\bf E}}\bdot\left\{{\bf X},\;\fd{\cal G}{F_{\rm gc}} \right\}_{\rm gc} \nonumber \\
   &&+\;  (4\pi q)^{2} \int_{\bf Z} F_{\rm gc} \left(\frac{\delta^{\star}{\cal F}}{\delta{\bf E}}\bdot\left\{{\bf X},\; {\bf X}\right\}_{\rm gc}\bdot \frac{\delta^{\star}{\cal G}}{\delta{\bf E}}\right) \nonumber \\
    &&+\; 4\pi c  \int_{\bf X} \left(\fd{\cal F}{\bf E}\bdot\nabla\btimes\fd{\cal G}{\bf B} - \fd{\cal G}{\bf E}\bdot\nabla\btimes\fd{\cal F}{\bf B} \right),
\nonumber
 \end{eqnarray}
 where $\delta^{\star}(\cdots)/\delta{\bf E}$ is defined in Eq.~\eqref{eq:delta_star}. The guiding-center bracket \eqref{eq:gcVM_bracket} is analogous to the Vlasov-Maxwell bracket \eqref{eq:VM_bracket}, where the Maxwell sub-bracket (last term) is identical in both cases, while the guiding-center Vlasov sub-bracket (first term) is connected by guiding-center phase-space transformation of the Vlasov sub-bracket in Eq.~\eqref{eq:VM_bracket}. The Interaction sub-bracket in Eq.~\eqref{eq:VM_bracket}, proportional to $4\pi q$, is transformed into the guiding-center Interaction sub-brackets in Eq.~\eqref{eq:gcVM_bracket}, where the quadratic term involving the antisymmetric dyadic Poisson bracket $\{{\bf X},{\bf X}\}_{\rm gc}$ represents the moving electric-dipole contribution to guiding-center magnetization, which is absent in Eq.~\eqref{eq:VM_bracket}. We note that, in contrast to recent work of Burby \cite{Burby_2015,Burby_2017}, the guiding-center Vlasov-Maxwell bracket \eqref{eq:gcVM_bracket} is expressed in terms of functional derivatives involving the Vlasov-Maxwell fields $(F_{\rm gc},{\bf E},{\bf B})$ since there is no guiding-center polarization in our model.

\subsection{Guiding-center momentum conservation law}

The conservation laws of energy-momentum and angular momentum for the guiding-center Vlasov-Maxwell \eqref{eq:V_bracket}-\eqref{eq:E_bracket} were recently derived by Brizard and Tronci \cite{Brizard_Tronci_2016}. As an application of the guiding-center Vlasov-Maxwell bracket \eqref{eq:gcVM_bracket}, we explore the time derivative of the guiding-center Vlasov-Maxwell (vector-valued) momentum functional
\begin{equation}
{\cal P}_{\rm gc} \;\equiv\;  \int_{\bf X} {\bf P}_{\rm gc} \;=\;  \int_{\bf Z} F_{\rm gc}\;p_{\|}\bhat \;+\;  \int_{\bf X}\frac{{\bf E}\btimes{\bf B}}{4\pi\,c},
\label{eq:P_gcVM}
\end{equation}
where each component ${\cal P}_{\rm gc}^{z} \equiv  \int_{\bf X}{\bf P}_{\rm gc}\bdot\wh{\sf z}$ satisfies the functional evolution equation
 \begin{eqnarray}
 \pd{{\cal P}_{\rm gc}^{z}}{t} &=& \left[ {\cal P}_{\rm gc}^{z},\frac{}{} {\cal H}_{\rm gc}\right]_{\rm gc} \nonumber \\
  &=&  \int_{\bf Z} F_{\rm gc} \left\{ \fd{{\cal P}_{\rm gc}^{z}}{F_{\rm gc}},\; \fd{{\cal H}_{\rm gc}}{F_{\rm gc}} \right\}_{\rm gc} \nonumber \\
   &&-\; 4\pi q\, \int_{\bf Z} F_{\rm gc}\;\frac{\delta^{\star}{\cal P}_{\rm gc}^{z}}{\delta{\bf E}}\bdot\left\{{\bf X},\; \fd{{\cal H}_{\rm gc}}{F_{\rm gc}} \right\}_{\rm gc} \nonumber \\
   &&+\; 4\pi q \int_{\bf Z} F_{\rm gc} \frac{\delta^{\star}{\cal H}_{\rm gc}}{\delta{\bf E}}\bdot\left\{{\bf X}, \fd{{\cal P}_{\rm gc}^{z}}{F_{\rm gc}} \right\}_{\rm gc}  \nonumber \\
    &&+\;  (4\pi q)^{2}  \int_{\bf Z} F_{\rm gc} \frac{\delta^{\star}{\cal P}_{\rm gc}^{z}}{\delta{\bf E}}\bdot\left\{{\bf X},\; {\bf X}\right\}_{\rm gc}\bdot\frac{\delta^{\star}{\cal H}_{\rm gc}}{\delta{\bf E}}   \nonumber \\
   &&+\; 4\pi c  \int_{\bf X} \fd{{\cal H}_{\rm gc}}{{\bf B}}\bdot\nabla\btimes\fd{{\cal P}_{\rm gc}^{z}}{\bf E} \nonumber \\
    &&-\; 4\pi c  \int_{\bf X} \fd{{\cal P}_{\rm gc}^{z}}{{\bf B}}\bdot\nabla\btimes\fd{{\cal H}_{\rm gc}}{\bf E}.
  \label{eq:gcVM_Pgc}
 \end{eqnarray}
Here, the functional derivatives of the guiding-center Hamiltonian functional \eqref{eq:gc_H} are given in Eq.~\eqref{eq:delta_H}, and the functional derivatives of the $z$-component of the guiding-center Vlasov-Maxwell momentum \eqref{eq:P_gcVM} are
\begin{equation}
\left( \begin{array}{c}
\delta{\cal P}_{\rm gc}^{z}/\delta F_{\rm gc} \\
4\pi c\,\delta{\cal P}_{\rm gc}^{z}/\delta{\bf E} \\
4\pi c\,\delta{\cal P}_{\rm gc}^{z}/\delta{\bf B}
\end{array} \right) = \left( \begin{array}{c}
p_{\|}\,b_{z} \\
{\bf B}\btimes\wh{\sf z} \\
\wh{\sf z}\btimes{\bf E} + 4\pi q\int_{P}F_{\rm gc}\,\wh{\sf z}\bdot \mathbb{P}_{\|}
\end{array} \right),
\label{eq:delta_Pgc}
\end{equation}
and
\begin{eqnarray}
4\pi c\, \frac{\delta^{\star}{\cal P}_{\rm gc}^{z}}{\delta{\bf E}} & = & {\bf B}\btimes\wh{\sf z} \;+\; \mathbb{P}_{\|}\bdot\nabla\btimes\left({\bf B}\btimes\wh{\sf z}\right) \nonumber \\
 &=& {\bf B}^{*}\btimes\wh{\sf z} \;+\; (p_{\|}c/q)\;\nabla b_{z}.
 \end{eqnarray}
 In Eq.~\eqref{eq:gcVM_Pgc}, we now evaluate
\begin{eqnarray*}
4\pi q\; \frac{\delta^{\star}{\cal P}_{\rm gc}^{z}}{\delta{\bf E}}\bdot\left\{{\bf X}, K_{\rm gc}\right\}_{\rm gc} &=& \wh{\sf z}\bdot\nabla K_{\rm gc} + \left\{ \fd{{\cal P}_{\rm gc}^{z}}{F_{\rm gc}}, K_{\rm gc} \right\}_{\rm gc}, \\
4\pi q\; \frac{\delta^{\star}{\cal P}_{\rm gc}^{z}}{\delta{\bf E}}\bdot\left\{{\bf X},\; {\bf X}\right\}_{\rm gc}  &=& \wh{\sf z} \;-\; \left\{ {\bf X},\; \fd{{\cal P}_{\rm gc}^{z}}{F_{\rm gc}} \right\}_{\rm gc},
\end{eqnarray*}
so that 
\begin{eqnarray*} 
 &&4\pi q \frac{\delta^{\star}{\cal H}_{\rm gc}}{\delta{\bf E}}\bdot\left( \left\{{\bf X}, \fd{{\cal P}_{\rm gc}^{z}}{F_{\rm gc}} \right\}_{\rm gc} + 4\pi q\, \frac{\delta^{\star}{\cal P}_{\rm gc}^{z}}{\delta{\bf E}}\bdot\left\{{\bf X}, {\bf X}\right\}_{\rm gc} \right) \\
  &&=\; q\,\wh{\sf z}\bdot\left( {\bf E} \;+\frac{}{} \mathbb{P}_{\|}\bdot\nabla\btimes{\bf E}\right) \;\equiv\; q\,\wh{\sf z}\bdot{\bf E}^{*},
  \end{eqnarray*}
 and, hence, we find
 \begin{eqnarray}
 \pd{{\cal P}_{\rm gc}^{z}}{t} &=& \int_{\bf Z} F_{\rm gc} \left[ \wh{\sf z}\bdot\left( q\,{\bf E}^{*} \;-\frac{}{} \nabla K_{\rm gc} \right) \right] 
   \label{eq:gcVM_Pgc_2} \\
  &&+\; 4\pi c  \int_{\bf X} \fd{{\cal H}_{\rm gc}}{{\bf B}}\bdot\nabla\btimes\fd{{\cal P}_{\rm gc}^{z}}{\bf E} \nonumber \\
   &&-\; 4\pi c  \int_{\bf X} \fd{{\cal P}_{\rm gc}^{z}}{{\bf B}}\bdot\nabla\btimes\fd{{\cal H}_{\rm gc}}{\bf E}.
\nonumber
 \end{eqnarray}
Next, we find
\begin{eqnarray*}
4\pi c\;\fd{{\cal H}_{\rm gc}}{{\bf B}}\bdot\nabla\btimes\fd{{\cal P}_{\rm gc}^{z}}{\bf E} &=& \nabla\bdot\left(\frac{|{\bf B}|^{2}}{8\pi}\,\wh{\sf z}\right) + \int_{P} F_{\rm gc}\,\wh{\sf z}\bdot\nabla K_{\rm gc}, \\
4\pi c\;\fd{{\cal P}_{\rm gc}^{z}}{{\bf B}}\bdot\nabla\btimes\fd{{\cal H}_{\rm gc}}{\bf E} &=& -\;\nabla\bdot\left[\left(\frac{{\bf E}{\bf E}}{4\pi} \;-\; \frac{|{\bf E}|^{2}}{8\pi}\,{\bf I}\right)\bdot\wh{\sf z}\right] \nonumber \\
 &&+\; \int_{P} F_{\rm gc}\; q\,{\bf E}^{*}\bdot\wh{\sf z},
\end{eqnarray*}
where we made use of Eq.~\eqref{eq:gc_divE}, so that Eq.~\eqref{eq:gcVM_Pgc} yields the guiding-center Vlasov-Maxwell momentum conservation law $\partial{\cal P}_{\rm gc}/\partial t = [{\cal P}_{\rm gc}, {\cal H}_{\rm gc}]_{\rm gc} = 0$. The derivation of the angular guiding-center momentum conservation law $\partial{\cal P}_{{\rm gc}\varphi}/\partial t = [{\cal P}_{{\rm gc}\varphi}, {\cal H}_{\rm gc}]_{\rm gc} = 0$, where ${\cal P}_{{\rm gc}\varphi} =  \int_{\bf X} {\bf P}_{\rm gc}\bdot\partial{\bf X}/\partial\varphi$ follows similar steps.

\subsection{Guiding-center Casimir functionals}

Casimir functionals ${\cal C}$ satisfy the bracket property $[{\cal C},{\cal K}]_{\rm gc} = 0$, which holds for an arbitrary functional ${\cal K}$. A standard example is the guiding-center entropy functional (omitting Boltzmann's constant)
\begin{equation}
{\cal S}_{\rm gc}[F_{\rm gc},{\bf B}] \;\equiv\; -\;\int_{\bf Z} F_{\rm gc}\;\ln\left(F_{\rm gc}/\frac{}{}B_{\|}^{*}\right),
\label{eq:S_gc}
\end{equation}
for which we obtain
\begin{eqnarray}
\left[{\cal S}_{\rm gc},\frac{}{} {\cal K}\right]_{\rm gc} &=&  \int_{\bf Z} F_{\rm gc} \left\{ \fd{{\cal S}_{\rm gc}}{F_{\rm gc}},\; \fd{\cal K}{F_{\rm gc}} \right\}_{\rm gc} 
  \label{eq:gcS_bracket} \\
  &&+\; 4\pi q \int_{\bf Z} F_{\rm gc}\;\frac{\delta^{\star}{\cal K}}{\delta{\bf E}}\bdot\left\{{\bf X},\; \fd{{\cal S}_{\rm gc}}{F_{\rm gc}} \right\}_{\rm gc}  \nonumber \\
    &&-\; 4\pi c  \int_{\bf X} \fd{\cal K}{\bf E}\bdot\nabla\btimes\fd{{\cal S}_{\rm gc}}{\bf B},
\nonumber
 \end{eqnarray}
 where $\delta{\cal S}_{\rm gc}/\delta F_{\rm gc} \;=\; -1 - \ln(F)$, with $F \equiv F_{\rm gc}/B_{\|}^{*}$, and
 \[ \fd{{\cal S}_{\rm gc}}{\bf B} \;=\; \int_{P} \left[ F\;\bhat \;-\; \frac{q}{c}\,\mathbb{P}_{\|}\bdot\left({\bf B}^{*}\;\pd{F}{p_{\|}} \;+\; \frac{c\bhat}{q}\btimes\nabla F \right) \right], \]
which is obtained after using the magnetic variations
 \begin{equation}
 \left. \begin{array}{rcl}
 \delta{\bf B}^{*} &=& \delta{\bf B} + \nabla\btimes(\mathbb{P}_{\|}\bdot\delta{\bf B}) \\
 (c/q)\delta\bhat &=& \delta{\bf B}\bdot\partial\mathbb{P}_{\|}/\partial p_{\|} \\
 \delta B_{\|}^{*} &=& \delta\bhat\bdot{\bf B}^{*} + \bhat\bdot\delta{\bf B}^{*}
 \end{array} \right\}.
 \label{eq:delta_B}
 \end{equation}
 Hence, we find the guiding-center bracket identity
\begin{eqnarray*}
\left[{\cal S}_{\rm gc},\frac{}{} {\cal K}\right]_{\rm gc} &=& -\;\int_{\bf Z} B_{\|}^{*} \left\{ F,\; \fd{\cal K}{F_{\rm gc}} \right\}_{\rm gc}  \\
  &&-\; 4\pi q \int_{\bf Z} \frac{\delta^{\star}{\cal K}}{\delta{\bf E}}\bdot\left({\bf B}^{*}\;\pd{F}{p_{\|}} \;+\; \frac{c\bhat}{q}\btimes\nabla F \right)   \\
    &&-\; 4\pi c  \int_{\bf X} \fd{{\cal S}_{\rm gc}}{\bf B}\bdot\nabla\btimes\fd{\cal K}{\bf E} \;\equiv\; 0,
 \end{eqnarray*}
 where the first term on the right side vanishes since it is an exact phase-space divergence, while the last term cancels out the second term. The expression \eqref{eq:S_gc} for the guiding-center Vlasov-Maxwell entropy was recently mentioned by Burby and Tronci \cite{Burby_Tronci_2017} and might find applications in the dissipative guiding-center bracket formulation \cite{Kaufman_1984,Morrison_1984,Morrison_1986} of the guiding-center Vlasov-Maxwell-Landau model (e.g., see Ref.~\cite{Iorio_Hirvijoki_2021}).
 
\subsection{Jacobi property of the guiding-center Vlasov-Maxwell bracket}

We now verify that the guiding-center bracket \eqref{eq:gcVM_bracket} satisfies the Jacobi property \eqref{eq:Jacobi}. According to the Bracket theorem \cite{Morrison_2013},  the proof of the Jacobi property involves only the explicit dependence of the guiding-center Vlasov-Maxwell bracket \eqref{eq:gcVM_bracket} on the guiding-center fields ${\sf J}_{\rm gc}^{ab}(F_{\rm gc}, {\bf B})$, where we note that the dependence on the magnetic field ${\bf B}$ enters through the guiding-center Poisson bracket \eqref{eq:PB_gc}, while the electric field ${\bf E}$ is explicitly absent. 

Hence, we can write the double-bracket involving three arbitrary guiding-center functionals $({\cal F}, {\cal G}, {\cal K})$:
\begin{eqnarray}
\left[[{\cal F},{\cal G}]_{\rm gc},\frac{}{} {\cal K}\right]_{\rm gc}^{P} &=& \int_{\bf Z} F_{\rm gc} \left\{ \frac{\delta^{P}[{\cal F}, {\cal G}]_{\rm gc}}{\delta F_{\rm gc}}, \fd{\cal K}{F_{\rm gc}} \right\}_{\rm gc} \nonumber \\
 &&+ 4\pi q \int_{\bf Z} F_{\rm gc}\,\frac{\delta^{\star}{\cal K}}{\delta{\bf E}}\vb{\cdot}\left\{ {\bf X},  \frac{\delta^{P}[{\cal F}, {\cal G}]_{\rm gc}}{\delta F_{\rm gc}} \right\}_{\rm gc} \nonumber \\
 &&- 4\pi c  \int_{\bf X}  \frac{\delta^{P}[{\cal F}, {\cal G}]_{\rm gc}}{\delta {\bf B}}\bdot\nabla\btimes\fd{\cal K}{\bf E},
 \label{eq:Jac_fg-k}
\end{eqnarray}
where the terms involving $\delta^{P}[{\cal F}, {\cal G}]_{\rm gc}/\delta{\bf E}$ vanish on the basis of the Bracket theorem. Here, the functional derivative $\delta^{P}[{\cal F}, {\cal G}]_{\rm gc}/\delta
F_{\rm gc}$ involves the explicit dependence on the guiding-center Vlasov distribution $F_{\rm gc}$ in the Vlasov and Interaction sub-brackets, while $\delta^{P}[{\cal F}, {\cal G}]_{\rm gc}/\delta{\bf B}$ in the Maxwell sub-bracket involves the explicit dependence  on the magnetic field ${\bf B}$, which appears through $(\bhat,{\bf B}^{*},B_{\|}^{*})$ in the guiding-center Poisson bracket \eqref{eq:PB_gc} and the dyadic tensor \eqref{eq:P_par}, where $(\delta\bhat,\delta{\bf B}^{*},\delta B_{\|}^{*})$ are given in Eq.~\eqref{eq:delta_B}. 

The proof of the Jacobi property for the guiding-center bracket \eqref{eq:gcVM_bracket} involves using several identities derived from the guiding-center Poisson bracket \eqref{eq:PB_gc} leading to an expansion in powers of $\epsilon \equiv 4\pi q$ up to third order:
\begin{widetext}
\begin{eqnarray}
{\cal Jac}[{\cal F},{\cal G},{\cal K}] &=&  \int_{\bf Z} F_{\rm gc}\left( \left\{ \{ f,\; g\}_{\rm gc},\frac{}{} k \right\}_{\rm gc} + \left\{ \{ g,\; k\}_{\rm gc},\frac{}{} f \right\}_{\rm gc}  + \left\{ \{ k,\; f\}_{\rm gc},\frac{}{} g \right\}_{\rm gc} \right) 
\label{eq:Jacobi_4piq} \\
 &&+\,  \epsilon \int_{\bf Z} F_{\rm gc} \left[ F_{i}^{\star} \left( \left\{ \{ X^{i},\; g\}_{\rm gc},\frac{}{} k \right\}_{\rm gc} + \left\{ \{ g,\; k\}_{\rm gc},\frac{}{} X^{i} \right\}_{\rm gc}  + \left\{ \{ k,\; X^{i}\}_{\rm gc},\frac{}{} g \right\}_{\rm gc} \right) \;+\; \leftturn \right]  \nonumber \\
 &&+\,  \epsilon^{2} \int_{\bf Z} F_{\rm gc} \left[ F_{i}^{\star}\,G_{j}^{\star} \left( \left\{ \{ X^{i},\; X^{j}\}_{\rm gc},\frac{}{} k \right\}_{\rm gc} + \left\{ \{ X^{j},\; k\}_{\rm gc},\frac{}{} X^{i} \right\}_{\rm gc}  + \left\{ \{ k,\; X^{i}\}_{\rm gc},
 \frac{}{} X^{j} \right\}_{\rm gc} \right) \;+\; \leftturn \right] \nonumber \\
  &&+\, \epsilon^{3} \int_{\bf Z} F_{\rm gc} \left[ F_{i}^{\star}\,G_{j}^{\star}\,K_{\ell}^{\star} \left( \left\{ \{ X^{i}, X^{j}\}_{\rm gc},\frac{}{} X^{\ell} \right\}_{\rm gc} + \left\{ \{ X^{j}, X^{\ell}\}_{\rm gc},\frac{}{} X^{i} \right\}_{\rm gc}  
  + \left\{ \{ X^{\ell}, X^{i}\}_{\rm gc},\frac{}{} X^{j} \right\}_{\rm gc} \right) \right], \nonumber
\end{eqnarray}
\end{widetext}
where $(f,g,k) \equiv (\delta{\cal F}/\delta F_{\rm gc}, \delta{\cal G}/\delta F_{\rm gc}, \delta{\cal K}/\delta F_{\rm gc})$ and $({\bf F}^{\star},{\bf G}^{\star},{\bf K}^{\star}) \equiv (\delta^{\star}{\cal F}/\delta {\bf E}, 
\delta^{\star}{\cal G}/\delta{\bf E}, \delta^{\star}{\cal K}/\delta  {\bf E})$, while all additional terms have cancelled out exactly (details of the proof are presented elsewhere \cite{Brizard_Jac_2021}). Here, summation over the repeated indices $(i,j,\ell)$ denoting the components of the vector fields $({\bf F}^{\star},{\bf G}^{\star},{\bf K}^{\star})$ is implied, and the symbol $\leftturn$ denotes cyclic permutations of the functionals $({\cal F},{\cal G},{\cal K})$. 

In Eq.~\eqref{eq:Jacobi_4piq}, it is clear that the Jacobi property of the guiding-center Vlasov-Maxwell bracket \eqref{eq:gcVM_bracket} is inherited from the Jacobi property \eqref{eq:gcPB_Jacobi} of the guiding-center Poisson bracket \eqref{eq:PB_gc}, since each term in Eq.~\eqref{eq:Jacobi_4piq} vanishes identically because of this latter property. Hence, the Jacobi property for the guiding-center Vlasov-Maxwell bracket \eqref{eq:gcVM_bracket} holds under the condition \eqref{eq:div_Bstar}.

\section{Discussion}

In the present paper, we derived the Hamiltonian structure of the guiding-center Vlasov-Maxwell equations introduced by Brizard and Tronci \cite{Brizard_Tronci_2016}. The associated guiding-center momentum and angular-momentum conservation laws were also presented in Hamiltonian form in terms of the guiding-center Vlasov-Maxwell bracket \eqref{eq:gcVM_bracket}. Since the guiding-center kinetic energy and Poisson bracket associated with the guiding-center Vlasov-Maxwell model considered here are both independent of the electric field, the effects of guiding-center polarization only appear through the moving electric-dipole contribution to the guiding-center magnetization \eqref{eq:M_def}. 

Future work will consider the inclusion of guiding-center polarization into our guiding-center Vlasov-Maxwell equations, as well as applications of the guiding-center Hamiltonian structure in gauge-free gyrokinetic Vlasov-Maxwell theory \cite{Burby_Brizard_2019,Brizard_2021}, where the perturbed electromagnetic fields $({\bf E}_{1},{\bf B}_{1})$ appear explicitly in the gyrocenter Lagrangian (e.g., see Ref.~\cite{Brizard_gyVM_2021}). We also plan to use the guiding-center Vlasov-Maxwell bracket \eqref{eq:gcVM_bracket} to explore extensions of Hamiltonian functional perturbation theory \cite{Brizard_Chandre_2020}.

\acknowledgments

The Author wishes to acknowledge useful discussions with C.~Tronci and J.W.~Burby. This work was supported by the National Science Foundation grant No.~PHY-1805164.

\vspace*{0.1in}

\begin{center}
{\bf Data Availability Statement}
\end{center}

Data sharing is not applicable to this article as no new data were created or analyzed in this study.

\end{document}